%% file: main.tex
\documentclass[12pt]{article}

\input{preamble}

\title{\papertitle}
\author[1]{Caoilte \'O Ciardha}
\author[2]{Joel Scanlan}
\author[3]{Tegan Insoll}
\author[4]{Juha Nurmi}
\author[3]{Nina Vaaranen-Valkonen}

\affil[1]{University of Kent, Canterbury, UK}
\affil[2]{CSAM Deterrence Centre, Hobart, Australia}
\affil[3]{Protect Children, Helsinki, Finland}
\affil[4]{Tampere University, Tampere, Finland}
\date{2026}

\begin{document}

\maketitle

\input{sections/00_abstract}
\input{sections/01_introduction}

\input{sections/02_methods}
\input{sections/03_results}

\input{sections/04_discussion}

\input{sections/05_declarations}

\printbibliography

\end{document}

%% file: preamble.tex
\usepackage{graphicx}

\usepackage[a4paper,margin=1in]{geometry}

\newcommand{\papertitle}{Warning Message Content Increases Help Seeking in a Large-Scale Dark Web CSAM Intervention}

\usepackage{authblk}

\newif\ifcomments
\commentstrue   

\usepackage{xcolor}

\ifcomments
  \newcommand{\caoilte}[1]{\textcolor{blue}{[caoilte: #1]}}
  \newcommand{\joel}[1]{\textcolor{red}{[joel: #1]}}
  \newcommand{\tegan}[1]{\textcolor{teal}{[tegan: #1]}}
  \newcommand{\juha}[1]{\textcolor{violet}{[juha: #1]}}
  \newcommand{\nina}[1]{\textcolor{green}{[nina: #1]}}
\else
  \newcommand{\caoilte}[1]{}
  \newcommand{\joel}[1]{}
  \newcommand{\tegan}[1]{}
  \newcommand{\juha}[1]{}
  \newcommand{\nina}[1]{}
\fi

\usepackage{csquotes}
\usepackage[
  style=apa,
  backend=biber,
  natbib=true,
  doi=true,
  url=true
]{biblatex}

\providecommand{\noopsort}[1]{}

\usepackage{booktabs}

\usepackage{caption}
\usepackage{setspace}

\usepackage{hyperref}
\usepackage[nameinlink]{cleveref}

\hypersetup{
    pdftitle={\papertitle},
    pdfauthor={Caoilte \'O Ciardha, Joel Scanlan, Tegan Insoll, Juha Nurmi, Nina Vaaranen-Valkonen}
}

\input{Results/05_random_day}

\input{Results/01_click_daytype}

\input{Results/02_click_its}
\input{Results/03_recid_daytype}

\input{Results/04_recid_its}

\input{Results/08_descriptives}

%% file: Results/05_random_day.tex
\newcommand{\RandomOmnibusChiSq}{520.40}
\newcommand{\RandomOmnibusDf}{8}
\newcommand{\RandomOmnibusP}{<0.001}

\newcommand{\RandomNeutralCTR}{12.17}

\newcommand{\RandomharmnegativeOR}{1.76}
\newcommand{\RandomharmnegativeCILow}{1.66}
\newcommand{\RandomharmnegativeCIHigh}{1.86}

\newcommand{\RandomdistresspositiveOR}{1.12}
\newcommand{\RandomdistresspositiveCILow}{1.06}
\newcommand{\RandomdistresspositiveCIHigh}{1.19}

%% file: Results/01_click_daytype.tex
\newcommand{\ClickDaytypeOmnibusChiSq}{309.47}
\newcommand{\ClickDaytypeOmnibusDf}{9}
\newcommand{\ClickDaytypeOmnibusP}{<0.001}

\newcommand{\ClickharmnegativeOR}{1.60}
\newcommand{\ClickharmnegativeCILow}{1.34}
\newcommand{\ClickharmnegativeCIHigh}{1.90}

\newcommand{\ClickdistresspositiveOR}{1.11}
\newcommand{\ClickdistresspositiveCILow}{1.00}
\newcommand{\ClickdistresspositiveCIHigh}{1.22}

%% file: Results/02_click_its.tex
\newcommand{\ClickITSPreSlopeOR}{1.00}
\newcommand{\ClickITSPreSlopeCILow}{1.00}
\newcommand{\ClickITSPreSlopeCIHigh}{1.00}
\newcommand{\ClickITSPreSlopeP}{0.646}
\newcommand{\ClickITSLevelOR}{2.01}
\newcommand{\ClickITSLevelCILow}{1.74}
\newcommand{\ClickITSLevelCIHigh}{2.31}
\newcommand{\ClickITSLevelP}{<0.001}
\newcommand{\ClickITSSlopeChangeOR}{1.00}
\newcommand{\ClickITSSlopeChangeCILow}{0.99}
\newcommand{\ClickITSSlopeChangeCIHigh}{1.00}
\newcommand{\ClickITSSlopeChangeP}{0.165}

%% file: Results/03_recid_daytype.tex
\newcommand{\RecidDaytypeOmnibusChiSq}{5.70}
\newcommand{\RecidDaytypeOmnibusDf}{9}
\newcommand{\RecidDaytypeOmnibusP}{0.769}

\newcommand{\RecidharmnegativeOR}{1.05}
\newcommand{\RecidharmnegativeCILow}{0.84}
\newcommand{\RecidharmnegativeCIHigh}{1.30}

\newcommand{\RecidrandomdayOR}{0.94}
\newcommand{\RecidrandomdayCILow}{0.85}
\newcommand{\RecidrandomdayCIHigh}{1.04}

%% file: Results/04_recid_its.tex
\newcommand{\RecidITSLevelOR}{0.92}
\newcommand{\RecidITSLevelCILow}{0.75}
\newcommand{\RecidITSLevelCIHigh}{1.13}
\newcommand{\RecidITSLevelP}{0.417}
\newcommand{\RecidITSSlopeChangeOR}{1.00}
\newcommand{\RecidITSSlopeChangeCILow}{0.99}
\newcommand{\RecidITSSlopeChangeCIHigh}{1.00}
\newcommand{\RecidITSSlopeChangeP}{0.316}

%% file: Results/08_descriptives.tex
\newcommand{\DescrTotalSearchesOverall}{19,611,744}
\newcommand{\DescrTriggeringSearchesOverall}{3,105,800}
\newcommand{\DescrTotalClicksOverall}{386,497}
\newcommand{\DescrTriggeringRateOverall}{15.84\%}
\newcommand{\DescrCtrOverall}{12.44\%}

\newcommand{\DescrCtrPre}{8.73\%}
\newcommand{\DescrCtrPost}{15.67\%}

\newcommand{\DescrTotalSearchesCampaign}{10,491,634}
\newcommand{\DescrTriggeringSearchesCampaign}{1,661,882}

\newcommand{\DescrCtrCampaign}{15.67\%}

\newcommand{\DescrTotalSearchesRandom}{1,078,552}
\newcommand{\DescrTriggeringSearchesRandom}{162,873}

\newcommand{\DescrCtrRandom}{15.58\%}

%% file: sections/00_abstract.tex
\begin{abstract}
\noindent
Warning messages have been used to disrupt individuals seeking online child sexual abuse material (CSAM) and promote engagement with support services, yet large-scale field evidence on message content remains limited, particularly in high-anonymity environments. This study reports a field experiment on Ahmia.fi, a Tor search engine, examining how warning message content influences behavior. Across a 140-day period, almost 20 million searches were observed, with over 3 million searches containing known CSAM-related terms that triggered a warning linking to an anonymous self-help program. Users were exposed to warning messages varying in thematic content and framing, or a neutral message. Across a randomized comparison, a campaign-wide analysis, and interrupted time series models, message content consistently influenced engagement with help resources. All active messages increased click-through rates to help resources relative to the neutral condition, with a harm-focused message producing the strongest effects. At the platform level, click-through rates increased from \DescrCtrPre{} before the intervention to \DescrCtrPost{} during the campaign. These findings highlight the importance of message content in shaping responses to warning interventions, supporting an approach in which messaging is refined and adapted to increase engagement with support resources.

\end{abstract}


%% file: sections/01_introduction.tex
\section*{Introduction}

Digital networks have expanded the accessibility and distribution of child sexual abuse material (CSAM; \citealp{carr2017brief}). Estimates indicate that more than 300 million children experience technology-facilitated sexual exploitation and abuse annually \citep{childlight2024intothelight}. Data from the National Center for Missing and Exploited Children (NCMEC) demonstrate consistent high levels of offending across online platforms \citep{ncmec2024cybertipline}, including on mainstream platforms commonly used by the broader community.

The scale and growth of offending is evident in the accessibility of CSAM across the clear web including search engines \citep{steel2015web, westlake2017assessing}, legal pornography websites \citep{fortin2018online, morgan2018understanding, ray2014correlates}, peer-to-peer networks \citep{prichard2011internet, wolak2014measuring}, messaging platforms, and social media \citep{teunissen2022child}. The dark web, which operates as a privacy-focused environment using The Onion Router (Tor) network, enables a substantial volume of offending, with numerous sites dedicated to the distribution of CSAM \citep{Nurmi2024}. Offending on the dark web is linked to the clear web, acting as a coordination hub that drives demand for real-time exploitation on mainstream platforms \citep{drejer2024livestreaming, rajamaki2022osint}.

CSAM consumption is increasingly conceptualized as a preventable public health problem, requiring interventions across offending pathways to mitigate its harm \citep{letourneau2014need, wortley2012internet}. Although offline approaches to crime prevention are limited in their ability to reach users in the moment of risk, digital interventions provide a scalable mechanism to interrupt individuals actively seeking CSAM \citep{wortley2012internet,watters2026} and divert \textit{some} of them to sources of help. Search queries of problematic terms, attempts to access known CSAM URLs, and hash matching of known abuse images and videos provide detection points for intervention \citep{hunn2023implement}.

Many online intervention strategies---such as warning messages on social media or search platforms---direct at-risk individuals toward appropriate therapeutic support. Empirical evidence indicates that a considerable proportion of individuals who consume CSAM express a willingness to change their behavior \citep{insoll2024factors}. These individuals often face structural and psychological barriers to seeking help, including stigma and fear of legal repercussions \citep{levenson2017, swaby2023}. Established prevention programs and anonymous helplines have demonstrated effectiveness in reducing risk \citep{latth2022}, however, engagement rates remain low. A significant gap therefore remains between high-risk online activity and the uptake of therapeutic resources designed to facilitate desistance. Even small improvements to warning messages---when implemented at scale---have the potential to help close that gap by increasing engagement with support resources at critical moments of risk and, in doing so, contribute to the prevention of online and offline child sexual abuse.

This study presents a field experiment conducted on Ahmia.fi, a dark web search engine. Across a 140-day observation period (including a 70-day pre-intervention baseline and a 70-day intervention period), over 3 million search queries triggered a warning message. Users entering high-risk search terms were presented with one of nine warning message variants. Each warning message provided a direct link to the ReDirection program, an anonymous online self-help resource. By analyzing click-through rates and CSAM search volume, we systematically evaluated whether and how message content influences help-seeking behavior. This study's findings provide empirical evidence to inform the design of warning messages to enable prevention pathways for individuals at risk of CSAM offending.

\section*{Background}

Warning messages are a common behavioral tool to discourage individuals from harmful or risky behaviors. In a crime prevention context, they are often conceptualized as a situational crime prevention strategy intended to disrupt offending behavior \citep{mayhew1975crime, clarke2017situational, watters2026}. The aim is to modify the environment to increase the perceived effort and risk associated with an action, thereby deterring potential offenders at the point of decision-making \citep{mayhew1975crime}. They provide an opportunity to \textit{disrupt} criminal behavior as well as \textit{divert} would-be offenders toward safer or supportive alternatives.

The concept of warning messages originates in the physical environment, where interventions have been widely implemented across occupational safety and public health domains \citep{wogalter2021warnings}. Warnings, particularly those designed to convey risk or avoid hazards are most effective when they capture attention, clearly articulate the hazard, and provide explicit instructions for avoidance \citep{lenorovitz2012ratings, wogalter2018chip}. These principles have been adapted for online environments, ranging from passive advisory text to dialog boxes that require active user acknowledgment \citep{egelman2008warned, kaiser2021adapting}. 

In online safety contexts, warning messages have been deployed on the demand side to mitigate CSAM-seeking behavior \citep{ProtectChildren2025}. While efforts to address CSAM have historically prioritized supply-side content moderation, including techniques such as hash matching to identify known illegal material and blocking access to it, the scale and growth of the CSAM ecosystem have necessitated complementary, scalable approaches that target user behavior directly \citep{hunn2023implement}. The internet allows the immediate satisfaction of passing curiosities, legal or otherwise, with a low perceived risk of detection \citep{quayle2012organisational, wortley2012internet}. Automated warnings counteract this thought process by disrupting behavior, notifying users of risks and harms, and prompting reflection. While such messaging has been deployed on multiple online platforms, there is a scarcity of large-scale evaluations to establish the most effective methods \citep{Priceetal2024}.

Warning message use has expanded to large-scale operational application by major electronic service providers, including search engines, social media, file-sharing services, and media providers \citep{ProtectChildren2025}. 
In addition to blocking content that violates policies, search engines deploy prominent warnings and help-seeking resources at the top of results pages for CSAM-related queries, an approach associated with measurable reductions in offending behavior \citep{steel2015web
}. Despite many platforms deploying warning messaging, most have not publicly evaluated the warning text or approach \citep{ProtectChildren2025}. In one study to do so, a warning message and an interactive chatbot called ‘reThink’ was evaluated on Pornhub in the United Kingdom, triggered by CSAM-related search queries \citep{scanlan2026}. In the reThink evaluation, 82\% of the 2.8 million users intercepted did not submit additional CSAM queries during the session, and there were active referrals to support services. There was also a downward trend in searches for CSAM during the evaluation. In the second large scale study on the implementation of warning messages, a multi-country change in the format and wording of the Google \textit{OneBox} warning, triggered following searches that are classified as potentially CSAM seeking, was associated with a reduction in retriggering of the warning within search sessions \citep{Umbach2026DeterringCSAM}. This was further associated with an increase in clicks on a Seek Help button examined specifically in the UK, whose warning already included the button, which was being included in other jurisdictions. Significant increases in helpline traffic across five regions (Australia, Canada, Germany, Ireland, and the United Kingdom) were attributable to the OneBox launch.     

These findings provide evidence of a change in CSAM seekers' behavior following introduction of or updates to warning messages, particularly in terms of reduction in further CSAM seeking on that platform. However, empirical evidence comparing different warning message types remains limited, with a small number of experimental studies examining their effects. A \textit{honeypot} methodology---where users encounter a realistic but researcher-controlled opportunity to access illicit material---demonstrated the immediate deterrent effects of these interventions \citep{Scanlanetal2022}. Data indicate that users exposed to legal warnings, particularly those referencing IP tracing and/or law enforcement, disengage from harmful behavior at significantly higher rates than in control groups \citep{prichard2022a}. In a dark web context, warnings about IP tracing and detection are likely less salient, as users understand this is far less likely in that environment. Messages featuring referral pathways to confidential therapeutic services also yielded comparable desistance, indicating that non-punitive framing can successfully redirect individuals seeking borderline or illicit content \citep{prichard2022b, prichard2024effect}. In a study where participants were asked to rank clear web search engine warning messages, results suggested that legality-focused messages are perceived as most effective for deterrence, while messages emphasizing accessibility of help and personal benefit are perceived as more conducive to help-seeking behaviors \citep{OCiardhaInPrep}.   

To date, there have been no large-scale field studies systematically examining the impact of search engine warning message content on engagement with help resources, and no large-scale studies on the impact of warning messages on the dark web. Drawing on current warning message practices, research on barriers and facilitators of help seeking \citep{Hammer2024, Levenson2019}, and empirical findings from both deterrence-focused field studies \citep{prichard2022a} and help-seeking focused self-report studies \citep{OCiardhaInPrep}, we tested four theoretically and practically grounded warning message themes. These were: (1) \textit{legality and consequences}, emphasizing risks of detection, punishment, and broader personal costs; (2) \textit{harm}, highlighting the impact of the behavior on victims; (3) \textit{behavioral control and self-efficacy}, emphasizing the individual’s capacity to change and the availability of support; and (4) \textit{psychological need or distress}, prompting reflection on the emotional or psychological drivers and consequences of the behavior.

In addition, we examined message valence as a cross-cutting feature, comparing negatively framed appeals (emphasizing risk, harm, or barriers) with positively framed appeals (emphasizing relief, support, or opportunities for change). This design allowed us to assess not only whether different message themes varied in effectiveness, but also whether their impact depended on how they were framed.   

\subsection*{Study Context -- the Tor network and Ahmia.fi}

Background and underpinning research for this study drew primarily on messaging work focusing on the clear web. The Tor network, often referred to as the dark web, is designed around privacy-preserving principles to enable communication and resist censorship. However, this censorship-resistant architecture also enables the distribution of substantial volumes of CSAM \citep{Guitton2013, Owen2016, Gannon2023, Nurmi2024}. CSAM-seeking behavior accounts for a disproportionate share of activity on the Tor network compared to other illicit content \citep{Spitters2014,Owen2016, Dalins2018}. This context underscores both the importance of effective intervention on the dark web and the possibility that warning messages operate through different mechanisms than those observed on the clear web.

A key implication of this environment is that the anonymity inherent in the Tor network fundamentally changes how warning-based interventions operate. On the clear web, warning messages can leverage deterrence mechanisms by highlighting law enforcement monitoring or IP tracing \citep{prichard2022a,Scanlanetal2022}. In contrast, such signals are less credible in the Tor environment, where IP tracing is not feasible and attributing activity to individual users remains challenging. These constraints do not eliminate the deterrent function of warning messages, which inherently interrupt behavior and signal that the activity is prohibited. However, they limit the credibility of more explicit enforcement-based deterrence cues (e.g., tracking or identification). As a result, while deterrence remains a core component of these interventions, this context suggests a greater emphasis on additional message features that facilitate reflection and engagement with anonymous therapeutic support \citep{letourneau2014need,wortley2012internet}. CSAM searching on the dark web may also reflect more deliberate and sustained engagement with offending behavior, involving individuals who have moved beyond more ambiguous or exploratory forms of searching. This may reduce the availability of (self-) justifications and shape how individuals respond to warning messages. 

Ahmia.fi serves as a search engine for locating Tor-hosted websites and is a popular entry point for users accessing the dark web \citep{Winter2018}. Established in 2014 with support from the Tor Project, Ahmia currently processes approximately 140,000 queries per day. Although Ahmia maintains user anonymity, recent self-report research on the platform identifies the CSAM-seeking population as predominantly male (76\%) and primarily composed of young adults aged 18 to 24 (45\%) and 25 to 34 (30\%), with an additional significant proportion under 18 \citep{protectchildren2026csam}.

Ahmia.fi has incorporated early-stage interventions into its search architecture. The search engine uses text mining and Bayesian classifiers to filter CSAM-related onion websites and actively blocks sexual search queries \citep{Nurmi2024, Nurmi2026}. Instead of returning empty results, these blocked queries serve as an intervention point, with warning messages that direct users to self-help programs and therapeutic resources. As a result of these interventions, between 2017 and 2024, Ahmia observed a 57.7\% decrease in CSAM searches, while the site's overall search volume doubled \citep{Nurmi2026}. The inaccessibility of CSAM via Ahmia.fi suggests that people searching for this content there have taken a step towards the anonymity afforded by the dark web but may not have yet developed sufficient expertise to effectively navigate that environment. This situates Ahmia.fi at a crucial intervention point along an offending trajectory. 

This study uses Ahmia's intervention infrastructure to systematically assess how different warning messages influence the rate at which users seek therapeutic help and the overall CSAM-seeking activity. This empirical evaluation of deterrence messaging in a live high-anonymity environment advances understanding of the efficacy of digital interventions.

%% file: sections/02_methods.tex
\section*{Method}

\subsection*{Design}

The study combined a randomized field experiment with a quasi-experimental campaign evaluation. The campaign ran on Ahmia.fi, a public search engine for onion websites. Prior to the campaign launch, searches using a banned term would trigger a warning message encouraging users to seek help via the ReDirection Self-Help Program. During the campaign period, banned searches instead triggered one of eight warning messages or a neutral message. The platform rotated through these nine fixed warning messages, with a randomized-warning day occurring every tenth day. Each day therefore corresponded to one of ten day-level conditions: nine fixed message conditions and a randomized-warning condition. The campaign ran for 70 days (1 September 2025 to 9 November 2025), ensuring that each day condition was presented once for every day of the week. Data were compared with the 70-day pre-campaign period (23 June 2025 - 31 August 2025). 

\subsection*{Banned Searches}
Ahmia.fi maintains a list of banned CSAM-related search terms. During the time period of this study, there were 1,281 terms listed. Terms are based on existing CSAM detection lists as well as monitoring of platform-specific search patterns. As a result, search terms that may reflect normative behavior on other platforms (e.g., \textit{porn}) are treated as high-risk indicators in this context, where broad sexual queries are strongly associated with CSAM-seeking \citep{Nurmi2024}, and therefore banned on this search engine.

\subsection*{Messages}
\subsubsection*{Pre campaign period}
Prior to the campaign launch, searchers triggering a warning were shown a static page informing users that help was available for individuals concerned about their use of child sexual abuse material or their sexual interest in children. The page presented several options, including links to (1) an anonymous questionnaire seeking to understand the experiences of people searching for sexual images or videos of children in order to develop more effective support resources, (2) the ReDirection Self-Help Program designed to help individuals manage urges to view sexual images or videos of children, including a Tor-accessible version of the site, and (3) Helplinks.eu, a directory of over 30 additional support services for individuals with a sexual interest in children. A screenshot of the pre-campaign warning page is provided in the Supplementary Materials.

The specific text promoting help seeking read: “Are you struggling with your use or urges to use sexual images or videos of children? Start the ReDirection program today and learn to take control of your thoughts and behavior. The self-help program is suitable for everyone, and it is secure, anonymous, and free.”

\subsubsection*{Campaign period}

\begin{table}[htbp]
\centering
\caption{Warning message text used in the campaign (English versions)}
\label{tab:warning_messages}
{\small
\input{Tables/warning_messages_table}
}
\end{table}

During the campaign period, searches triggering a warning were instead shown a page with a link to the ReDirection program first, with message content that varied by condition, and the same questionnaire as prior to the campaign launch below. Each of the messages linking to the ReDirection program retained common boilerplate elements but varied in persuasive content across four categories. The first category focused on legality and consequences, highlighting the legal risks and potential personal consequences associated with accessing CSAM. The second category emphasized the harm inflicted on children by viewing CSAM. The third category focused on the individual's ability to take control of their situation and access support. The fourth category focused on psychological distress, highlighting the potential emotional and psychological difficulties that may arise from viewing CSAM. Messages were presented in English and in Spanish.

\begin{figure}
    \centering
    \includegraphics[width=0.75\linewidth]{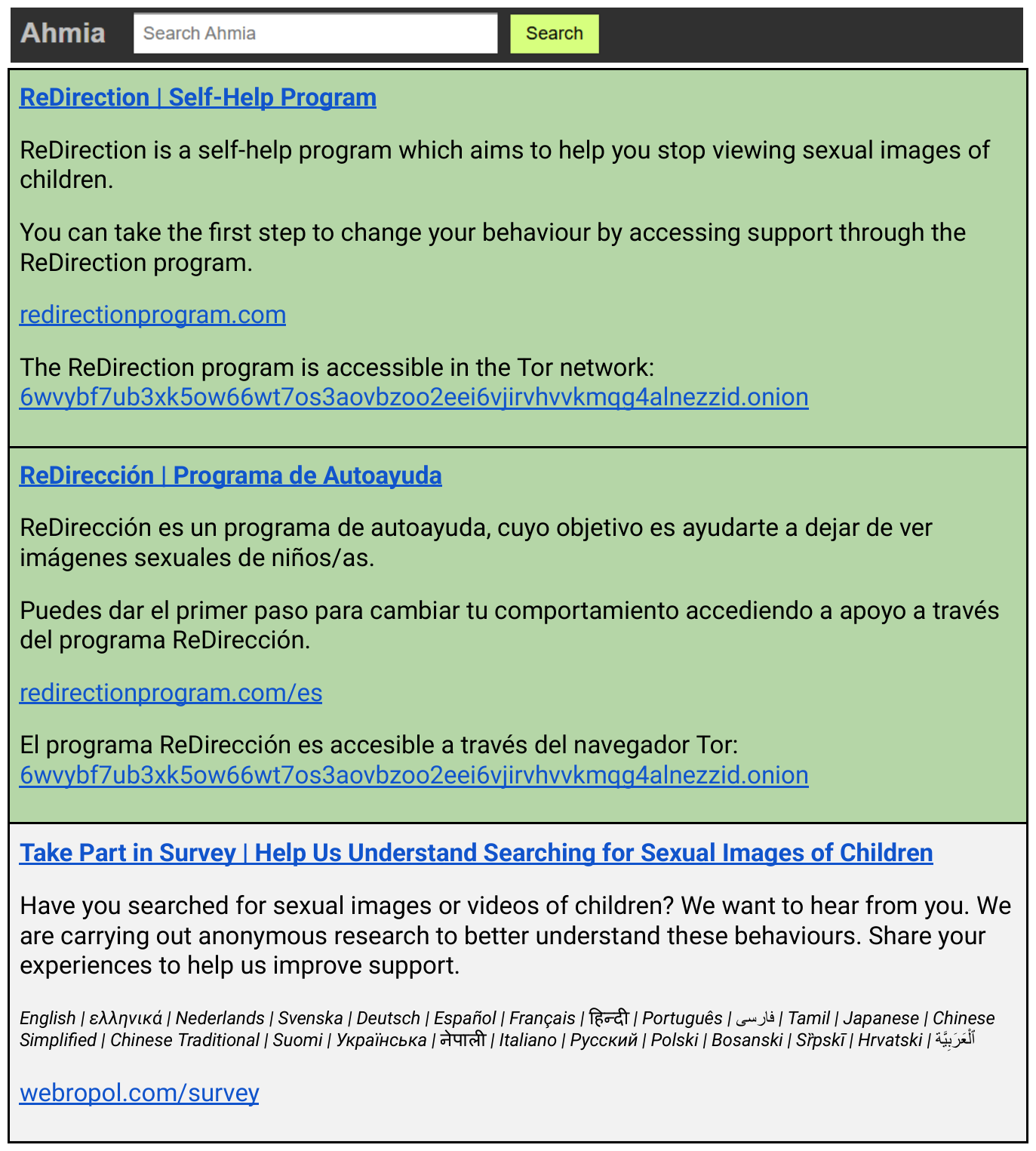}
    \caption{Mock-up of neutral warning message as it appeared to searchers following a banned search }
    \label{fig:campaign_message_neutral}
\end{figure}

For each thematic category, there were two versions of the message, one negatively framed and one positively framed. Negatively framed messages emphasized risks, harms, or barriers, whereas positively framed messages emphasized opportunities, accessibility of support, or the possibility of relief. The neutral message provided information about the self-help program: “ReDirection Self-Help Program. ReDirection is a self-help program which aims to help you stop viewing sexual images of children.” See Figure \ref{fig:campaign_message_neutral} for a mock-up of how the neutral warning message appeared to searchers following a banned search.

\subsection*{Data Sources}
Our data consisted of search queries on Ahmia.fi. For each day of the campaign and the preceding 70 days we had total search volume, number of searches triggering a warning, and the number of times a help link on the warning was clicked on. This was aggregated at the day level for searches on days where only one message type was triggered. On days where messages were served randomly, these data were further broken down by message type. The number of warning exposures represents the number of times a warning message was displayed in response to a search. Consistent with the privacy-preserving principles of the Tor network, Ahmia does not track user IP addresses, use persistent cookies, or de-anonymize traffic. Instead, it only retains minimal HTTP logs to support maintenance and behavioral research (\citealp{Nurmi2026}; see, e.g. \citealp{insoll2024factors, lahtinen2025investigating}). Because Ahmia.fi does not use persistent identifiers, the available data consist of aggregated counts of searches rather than search sessions by individual users.  

\subsection*{Analytic Strategy}
All analyses were conducted in R (version 4.5.2). Outcomes were analyzed using binomial logistic regression models applied to aggregated daily counts. Model coefficients and confidence intervals were based on Wald statistics, and omnibus tests of message condition were assessed using likelihood ratio $\chi^2$ tests.

\subsubsection*{Randomized-Day Analysis}
To estimate message effects under random assignment, we analyzed the subset of days on which message conditions were randomized. Click-through to help resources was modeled using logistic regression with the number of clicks treated as successes and the remaining warning exposures as non-clicks. Message condition was included as a categorical predictor with the neutral message as the reference category, and day fixed effects were included to compare message conditions within the same day while controlling for day-level variation.

\subsubsection*{Campaign-Period Analysis}
We estimated logistic regression models using all campaign-period data to examine (1) click-through and (2) warning message triggering across message conditions during the deployment period. The outcomes were (1) number of clicks relative to the number of warning exposures and (2) number of warning exposures relative to total searches. Models included message condition with neutral message as the reference category, day of week, and a linear time trend to account for temporal variation during the campaign.

To account for residual temporal dependence in the daily series, we estimated heteroskedasticity and autocorrelation consistent (Newey--West HAC; \citealp{NeweyWest1987}) standard errors with a lag length of 7 days, implemented using the \texttt{sandwich} and \texttt{lmtest} packages in R (\citealp{sandwich2020,lmtest2002}). For models showing significant focal effects, we report sensitivity analyses using alternative Newey–West lag specifications (4–14 days) in Supplementary Materials.

\subsubsection*{Interrupted Time Series (ITS) Analysis}
To examine changes associated with the deployment of the redesigned warnings on (1) CTR and (2) rate of warning-triggering searches, we estimated an interrupted time-series logistic regression model using aggregated daily counts from both the pre- and post-deployment periods (\citealp{Bernal2017}). These models included a linear time trend, an indicator for the post-deployment period, and a post-deployment slope term, along with day of week as a control variable. Inference again used Newey--West HAC standard errors with a 7-day lag, with lag-sensitivity analyses reported in Supplementary Materials for models showing significant focal effects.

\subsection*{Ethics}
The study received ethical approval from the Ethics Committee of the Tampere Region (Finland) on 16 June 2025 (Statement 68/2025).

%% file: Tables/warning_messages_table.tex
    \begin{tabular}{p{4.2cm} p{10cm}}
\toprule
\textbf{Message Category} & \textbf{Message Text} \\
\midrule

\textbf{Legality \newline Negative-framed} &
ReDirection \textbar\ \textbf{Child sexual abuse imagery is illegal.} Accessing sexual images of children puts you at risk of arrest and may cost you your relationships, your job, or your freedom. You can take the first step to change your behaviour by accessing support through the ReDirection program.\\
\\
\textbf{Legality \newline Positive-framed} &
ReDirection \textbar\ \textbf{Child sexual abuse imagery is illegal.} Getting professional help may reduce the risk of arrest and help you keep your relationships, your job, your freedom. You can take the first step to change your behaviour by accessing support through the ReDirection program.\\
\\
\textbf{Harm \newline Negative-framed} &
ReDirection \textbar\ \textbf{Child sexual abuse imagery causes harm to children.} Searching for and viewing sexual images of children adds to that harm. You can take the first step to change your behaviour by accessing support through the ReDirection program.\\
\\
\textbf{Harm \newline Positive-framed} &
ReDirection \textbar\ \textbf{Child sexual abuse imagery causes harm to children.} Getting help to stop viewing sexual images of children is one way you can stop that cycle of harm. You can take the first step to change your behaviour by accessing support through the ReDirection program.\\
\\
\textbf{Control \newline Negative-framed} &
ReDirection \textbar\ \textbf{Getting help for child sexual abuse imagery starts with a single click.} Don’t dwell on the barriers stopping you from getting anonymous help. You can take the first step to change your behaviour by accessing support through the ReDirection program.\\
\\
\textbf{Control \newline Positive-framed} &
ReDirection \textbar\ \textbf{Getting help for child sexual abuse imagery starts with a single click.} It’s easier than you think to get anonymous help to stop viewing sexual images of children. You can take the first step to change your behaviour by accessing support through the ReDirection program.\\
\\
\textbf{Distress \newline Negative-framed} &
ReDirection \textbar\ \textbf{How is searching for child sexual abuse imagery affecting you?} Take a moment to think about how searching for sexual images of children is likely to cause you feelings of shame, guilt, and anxiety. You can take the first step to change your behaviour by accessing support through the ReDirection program.\\
\\
\textbf{Distress \newline Positive-framed} &
ReDirection \textbar\ \textbf{How is searching for child sexual abuse imagery affecting you?} Getting help to stop searching for sexual images of children can take away your feelings of shame, guilt, and anxiety. You can take the first step to change your behaviour by accessing support through the ReDirection program.\\
\\
\bottomrule
\end{tabular}

%% file: sections/03_results.tex
\section*{Results} 

Over the 140 day period covered by the study there were \DescrTotalSearchesOverall{} searches on Ahmia.fi, of which \DescrTriggeringSearchesOverall{} (or \DescrTriggeringRateOverall{} of searches) contained banned CSAM-related terms which triggered a warning. These warnings resulted in \DescrTotalClicksOverall{} clicks to help resources (i.e., \DescrCtrOverall{} of warnings resulted in a click). 

\subsection*{Click-Through Rate (CTR)}

\subsubsection*{Randomized Days}

\begin{figure}[htbp]
    \centering
    \caption{Randomized-day message effects on click-through rates}
    \includegraphics[width=\linewidth]{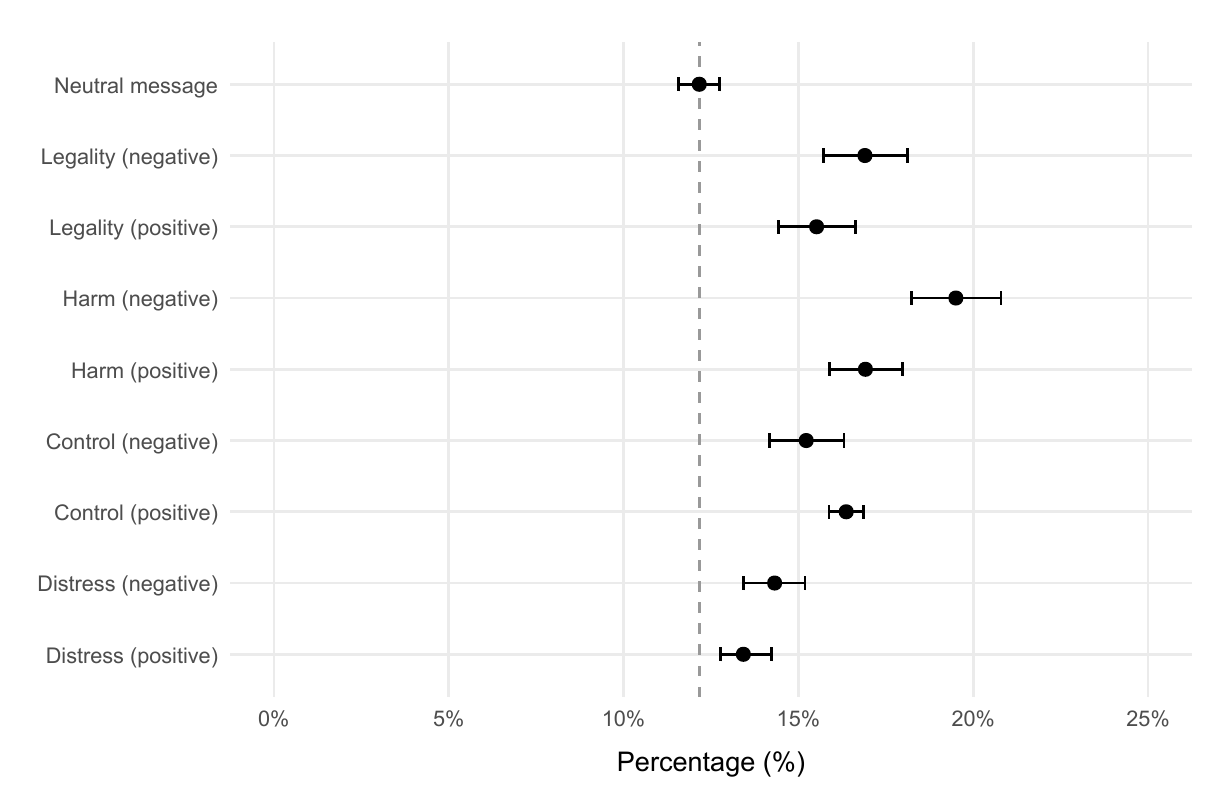}
    \begin{minipage}{0.9\linewidth}
    \footnotesize
    \textit{Note.} Points show the average observed click-through rate for each message condition across the randomized days. Error bars show 95\% confidence intervals obtained by bootstrapping across days. Dashed line allows comparison with the neutral condition.
\end{minipage}
\label{fig:random_day_pct}
\end{figure}

The campaign period included seven days on which warning messages were shown randomly. On these randomized days, there were \DescrTotalSearchesRandom{} searches, \DescrTriggeringSearchesRandom{} of which triggered a warning message. Across all conditions, the click-through rate (CTR) to a help resource was \DescrCtrRandom{}. Message condition significantly affected CTR in a binomial logistic regression with day fixed effects, 
$\chi^2(\RandomOmnibusDf) = \RandomOmnibusChiSq$, $p \RandomOmnibusP$. 
All message variants increased click through relative to the neutral message (all $p < .001$), 
with odds ratios ranging from \RandomdistresspositiveOR\ (95\% CI [\RandomdistresspositiveCILow, \RandomdistresspositiveCIHigh]) 
to \RandomharmnegativeOR\ (95\% CI [\RandomharmnegativeCILow, \RandomharmnegativeCIHigh]). 
The negatively-framed harm-focused messages produced the largest effects, followed by the positively-framed harm message and legality and control messages, 
with smaller but reliable effects for distress-focused messages. 
Consistent with these estimates, Figure \ref{fig:random_day_pct} displays the day-averaged CTRs for each message condition, including the neutral baseline. The neutral condition resulted in a CTR of \RandomNeutralCTR\%, compared with a pre-campaign baseline CTR of \DescrCtrPre, representing a 39.4\% relative increase.

\subsubsection*{Campaign Days}

We next examined how click through varied across the different day types used in the campaign, including the randomized-warning days. During this campaign period, there were \DescrTotalSearchesCampaign{} searches, \DescrTriggeringSearchesCampaign{} warning-triggering searches, and a CTR of \DescrCtrCampaign{}.

\begin{figure}[htbp]
    \centering
    \caption{Campaign period - message effects on click-through rates}
    \includegraphics[width=\linewidth]{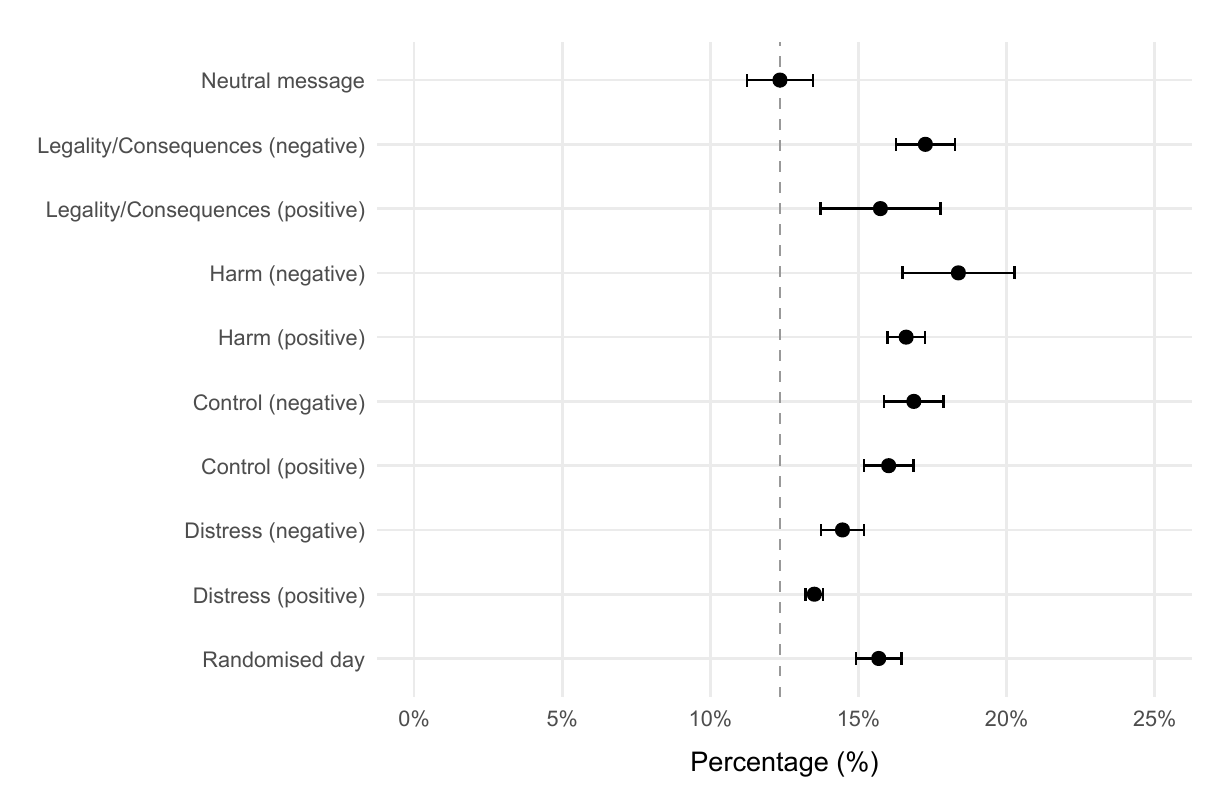}
    \begin{minipage}{0.9\linewidth}
    \footnotesize
    \textit{Note.} Points show model-predicted click-through rates for each message condition from a binomial logistic regression adjusting for weekday and a linear campaign time trend. Predicted probabilities are averaged across the observed campaign days. Error bars show 95\% confidence intervals based on Newey--West heteroskedasticity and autocorrelation consistent standard errors (lag = 7). Dashed line allows comparison with the neutral condition.
\end{minipage}
\label{fig:click_daytype}
\end{figure}

Message condition was associated with CTR in a binomial logistic regression adjusting for day of week and linear time trend with Newey--West HAC standard errors, $\chi^2(\ClickDaytypeOmnibusDf) = \ClickDaytypeOmnibusChiSq$, $p  \ClickDaytypeOmnibusP$. All HAC-based inferences were robust to alternative lag specifications
(4--14 days; see Supplementary Materials). 
All message conditions were associated with higher click through relative to the neutral message (all $p < .05$), 
with odds ratios ranging from \ClickdistresspositiveOR\ (95\% CI [\ClickdistresspositiveCILow, \ClickdistresspositiveCIHigh]) 
to \ClickharmnegativeOR\ (95\% CI [\ClickharmnegativeCILow, \ClickharmnegativeCIHigh]). 
The negatively-framed harm-focused messages again produced the largest effects, followed again by positively-framed harm and legality and control messages, 
with smaller effects for distress-focused messages.
Figure \ref{fig:click_daytype} shows the model-derived predicted CTRs for each message condition during the campaign period, including the neutral baseline. 

\subsubsection*{Interrupted Time Series (ITS)}

\begin{figure}[htbp]
    \centering
    \caption{Click-through rates over time with interrupted time-series trend}
    \includegraphics[width=\linewidth]{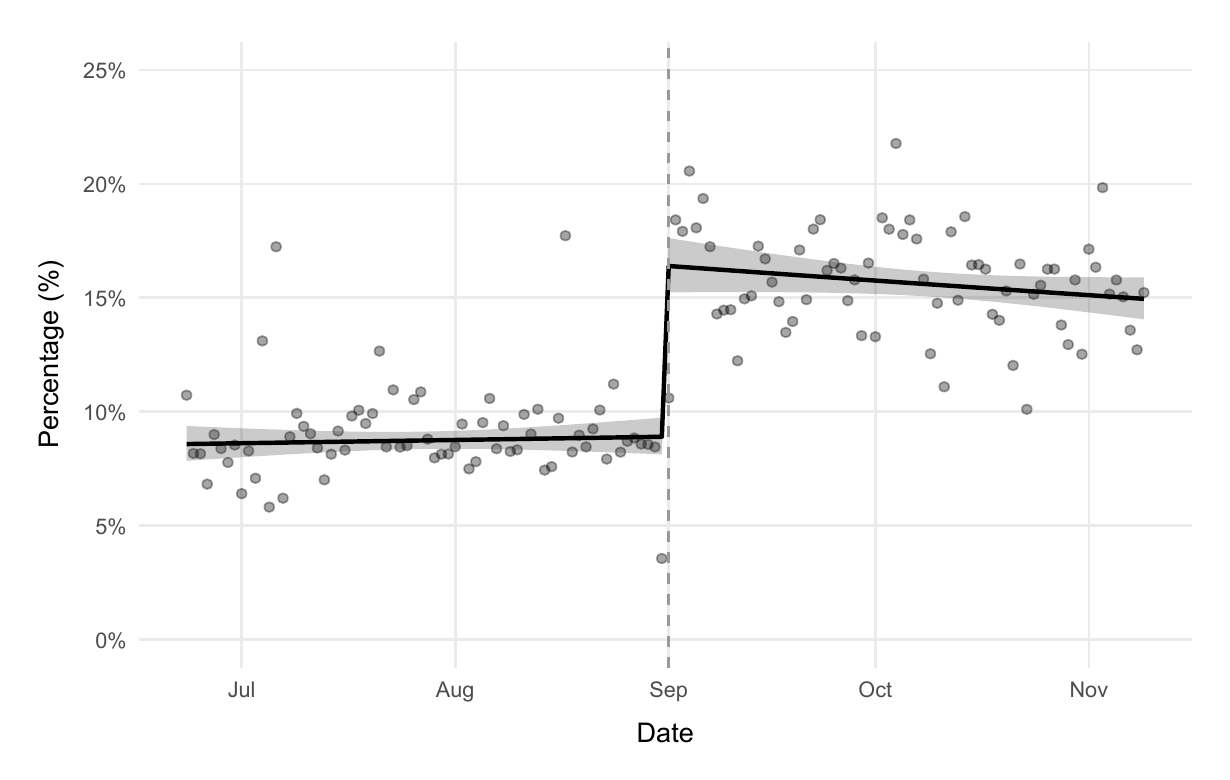}
    \begin{minipage}{0.9\linewidth}
    \footnotesize
    \textit{Note.} Lines show weekday-averaged model-predicted click-through rates from an interrupted time series logistic regression including a baseline time trend, an immediate post-campaign level change, and a post-campaign slope term. Shaded areas show 95\% confidence intervals based on Newey--West heteroskedasticity and autocorrelation consistent standard errors (lag = 7). Dashed line indicates the start of the campaign intervention.
\end{minipage}
\label{fig:click_its}
\end{figure}

We ran an interrupted time-series (ITS) analysis to examine the impact of the change in warning messages on Ahmia.fi at the campaign level. Prior to the campaign launch, CTR was  \DescrCtrPre{}, increasing to \DescrCtrPost{} in the campaign period. The ITS indicated a significant immediate increase in CTR at campaign launch, with an odds ratio of \ClickITSLevelOR\ (95\% CI [\ClickITSLevelCILow, \ClickITSLevelCIHigh], $p \ClickITSLevelP$). 
There was no clear evidence of a pre-campaign trend in CTR (\textit{OR} per day = \ClickITSPreSlopeOR, 95\% CI [\ClickITSPreSlopeCILow, \ClickITSPreSlopeCIHigh], $p = \ClickITSPreSlopeP$) or of a change in slope following campaign onset (\textit{OR} per day = \ClickITSSlopeChangeOR, 95\% CI [\ClickITSSlopeChangeCILow, \ClickITSSlopeChangeCIHigh], $p = \ClickITSSlopeChangeP$). Figure \ref{fig:click_its} shows the model-predicted CTR over time.

\subsection*{Warning-Trigger Rate}

We also examined whether the campaign affected the overall volume of searches triggering a warning message. Because session-level information was not available, these analyses assess changes in the aggregate warning-trigger rate rather than direct re-triggering by the same individual.

\subsubsection*{Campaign Days}

\begin{figure}[htbp]
    \centering
    \caption{Campaign period -- message effects on warning-trigger rate}
    \includegraphics[width=\linewidth]{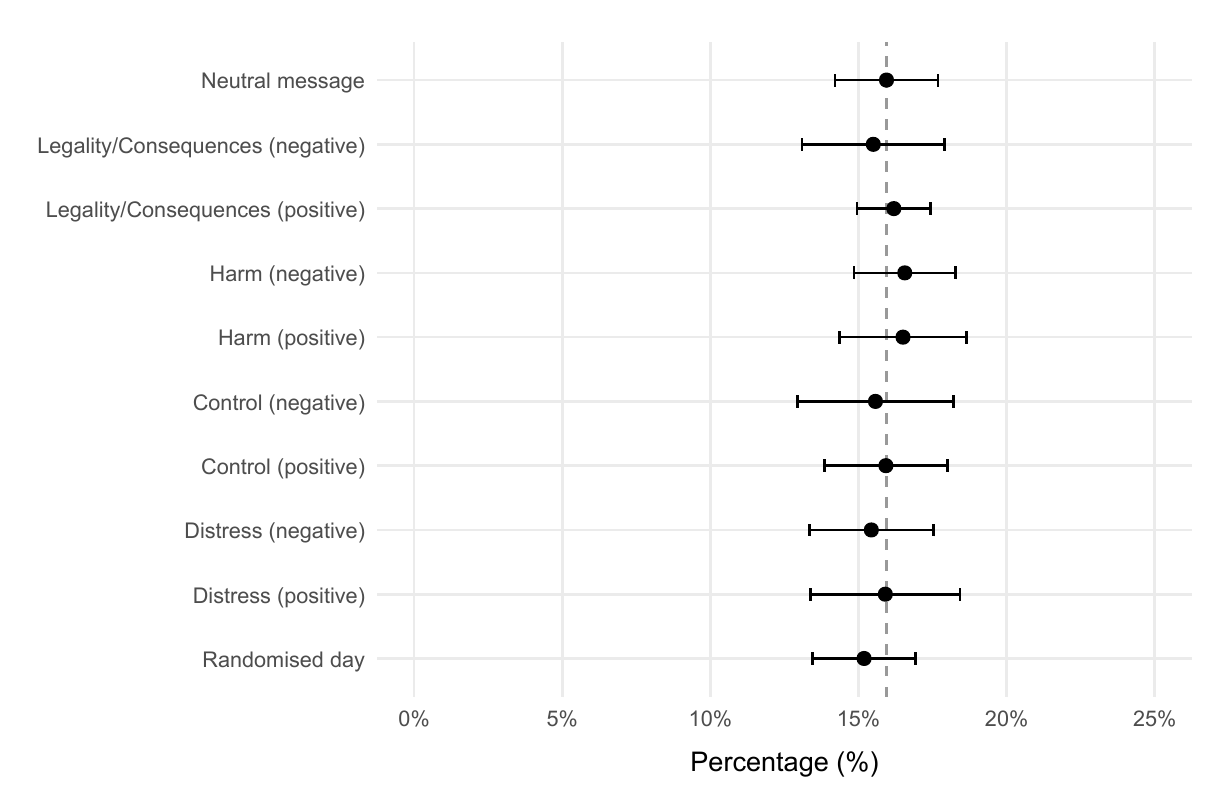}
    \begin{minipage}{0.9\linewidth}
    \footnotesize
    \textit{Note.} Points show model-predicted warning-trigger rates for each message condition from a binomial logistic regression adjusting for weekday and a linear campaign time trend. Predicted probabilities are averaged across the observed campaign days. Error bars show 95\% confidence intervals based on Newey--West heteroskedasticity and autocorrelation consistent standard errors (lag = 7). Dashed line allows comparison with the neutral condition.
\end{minipage}
\label{fig:recid_daytype}
\end{figure}

Message condition was not associated with warning-trigger rate in a binomial logistic regression adjusting for day of week and linear time trend, with Newey--West HAC standard errors, $\chi^2(\RecidDaytypeOmnibusDf) = \RecidDaytypeOmnibusChiSq$, $p = \RecidDaytypeOmnibusP$. 
No message condition differed reliably from the neutral message, 
with odds ratios ranging from \RecidrandomdayOR\ (95\% CI [\RecidrandomdayCILow, \RecidrandomdayCIHigh]) 
to \RecidharmnegativeOR\ (95\% CI [\RecidharmnegativeCILow, \RecidharmnegativeCIHigh]). 
Thus, unlike click through, the campaign-day analysis provided no clear evidence that message condition affected the rate of warning-triggering searches in either direction at the platform level. 
Figure \ref{fig:recid_daytype} shows the model-derived predicted warning-trigger rates for each message condition during the campaign period, including the neutral baseline.

\subsubsection*{Interrupted Time Series (ITS)}

\begin{figure}[htbp]
    \centering
    \caption{Warning-trigger rate over time with interrupted time-series trend}
    \includegraphics[width=\linewidth]{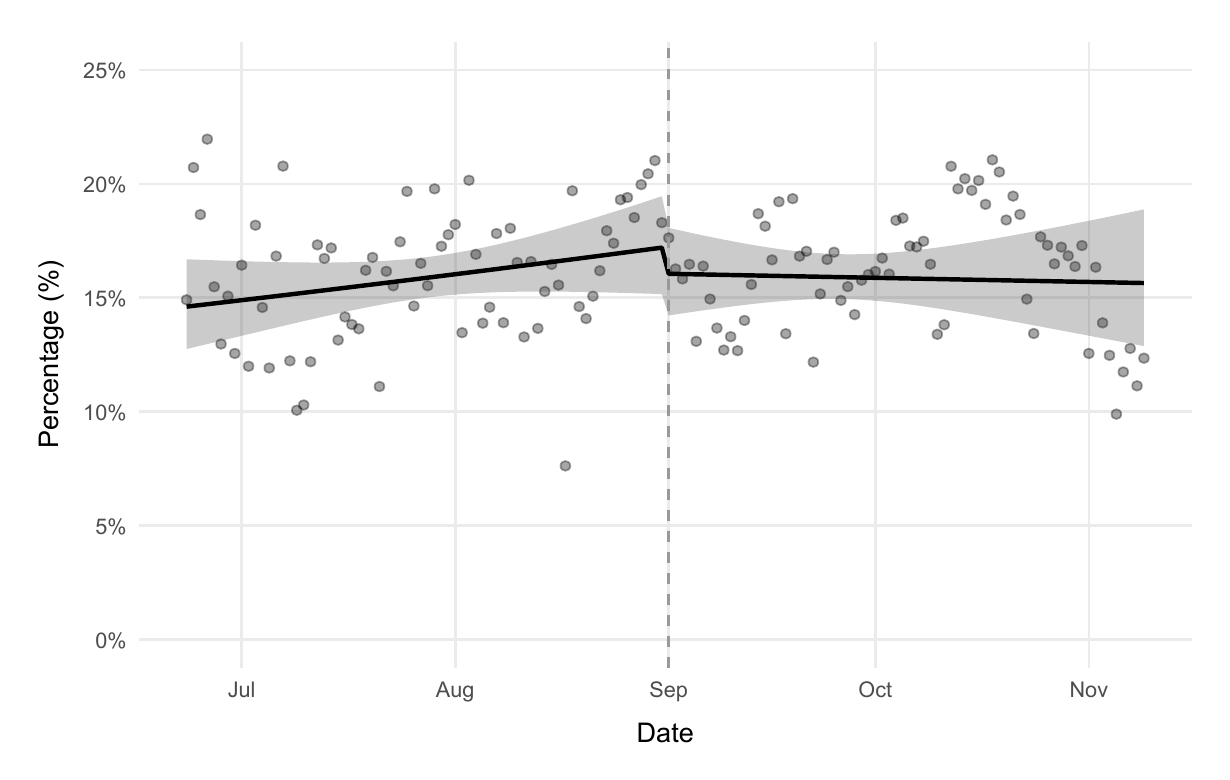}
    \begin{minipage}{0.9\linewidth}
    \footnotesize
    \textit{Note.} Lines show weekday-averaged model-predicted warning-trigger rates from an interrupted time series logistic regression including a baseline time trend, an immediate post-campaign level change, and a post-campaign slope term. Shaded areas show 95\% confidence intervals based on Newey--West heteroskedasticity and autocorrelation consistent standard errors (lag = 7). Dashed line indicates the start of the campaign intervention.
\end{minipage}
\label{fig:recid_its}
\end{figure}

An interrupted time series logistic regression examined changes in the warning-trigger rate at the start of the campaign, including a baseline time trend, an immediate post-campaign level change, and a post-campaign slope change, with Newey--West HAC standard errors. 
Again, there was no clear evidence of a change in warning-trigger rate at campaign launch (OR = \RecidITSLevelOR, 95\% CI [\RecidITSLevelCILow, \RecidITSLevelCIHigh], $p = \RecidITSLevelP$), 
nor evidence of a change in the post-campaign slope (OR per day = \RecidITSSlopeChangeOR, 95\% CI [\RecidITSSlopeChangeCILow, \RecidITSSlopeChangeCIHigh], $p = \RecidITSSlopeChangeP$). 
Figure \ref{fig:recid_its} shows the model-implied warning-trigger rate over time with the estimated campaign interruption.

%% file: sections/04_discussion.tex
\section*{Discussion}
This study examined the impact of redesigned warning messages in a large-scale field setting, using data from approximately 20 million searches on a Tor search engine, within which just over three million searches triggered warning messages. Our findings show that the content of warning messages served in response to banned searches had a measurable impact on click-through rates (CTR) to help resources. This was first demonstrated in the randomized-day experiment, in which warning message variants were randomly assigned, indicating that differences in CTR between messages were not attributable to external factors such as day of the week, seasonal effects, or contemporaneous events. All alternative warning messages significantly outperformed a neutral message. A similar pattern was observed across the 70-day campaign period, during which messages were rotated daily. In this analysis, all message conditions again showed higher CTR than days on which the neutral message was shown. Finally, when examined at the platform level using an interrupted time-series design, CTR increased from \DescrCtrPre{} prior to the campaign to \DescrCtrPost{} during the campaign period, corresponding to approximately a doubling in the odds of clicking through to help resources following the message redesign.

The deployment of the new warning messages was not associated with any detectable change in the rate of warning-triggering searches at the platform level. This suggests that the primary behavioral change resulting from the redesigned warnings was increased engagement with help resources rather than deterrence from searching (i.e., measurable reductions in warning-triggering searches). One possible explanation for the absence of detectable content-dependent deterrence effects is that baseline search volumes on Ahmia may already have been substantially reduced through earlier interventions, with a 57.7\% decline observed between 2017 and 2024 \citep{Nurmi2026}. In such contexts, there may be greater scope for gains in help seeking than for further reductions in search behavior, meaning that the impact of warning messages was more likely to be observed in increased engagement with support resources. 

We considered two possible alternative explanations for our omnibus findings of message impact on CTR. At campaign launch, the warning interface also underwent formatting changes. In the pre-campaign interface (see Supplementary Materials), users triggering a banned search were shown several stacked search-result style panels, including a survey invitation, links to the ReDirection program (both clear web and Tor versions), and an additional help resources page. In the redesigned interface used during the campaign, these elements were consolidated into a single warning page centered on the ReDirection program. The message was presented as a structured panel containing English and Spanish versions of the help-seeking text, with links to the program and the survey positioned below. This redesign simplified the interface and increased the visual prominence of the help-seeking message. These changes may account for some portion of the increase in CTR observed between the pre-campaign and campaign periods. The neutral message had a CTR of  \RandomNeutralCTR\% during random days, substantially higher than the pre-campaign baseline of \DescrCtrPre. This increase is unlikely to be attributable to message wording alone and provides a tentative estimate of the contribution of interface and presentation changes. However, interface changes do not explain the differences in CTR observed between message types during the campaign period or on randomized days. 

Additionally, the campaign rotated through message types in a fixed daily order. This was to ensure that every condition was presented once per day of the week. This may have introduced temporal confounds that were not fully captured by the inclusion of weekday and linear time covariates in the campaign-period models. Additionally, in this sequence, negatively framed messages were always followed by the positively-framed message on the same theme. However, this potential limitation did not apply to the randomized-day experiment, where message conditions were randomly displayed. Taken together, the randomized-day experiment and the consistent patterns observed in the campaign-period and interrupted time-series analyses provide convergent evidence that warning message content can influence engagement with help resources.

\subsection*{Message themes and valence}
Across both the randomized-day experiment and the broader campaign analysis, all active messages outperformed the neutral message, indicating that message content meaningfully influenced engagement with support resources. Within this overall pattern, negatively framed harm messages produced the highest CTR, with legality- and control-themed messages showing intermediate effects, and distress-themed messages the smallest effects among the active conditions.

The observed pattern differs in important ways from some prior research. In self-report work, behavior control- and distress-focused messages were perceived as most effective for promoting help seeking, whereas harm and legality messages are rated more highly for deterrence \citep{OCiardhaInPrep}. A similar distinction is reflected in a honeypot study examining attempts to access so-called \textit{barely legal} pornography \citep{prichard2022a}. There, deterrence-oriented messages---particularly those emphasizing illegality---were associated with the largest reductions in access, with harm-focused messages (including harm to the viewer in the form of long-term distress, which aligns with the distress category framing used here) showing smaller effects that were not statistically reliable. However, in real-world settings, the distinction between disruption and diversion may be less sharply resolved. One plausible explanation is that these warning messages function in part as situational interruptions rather than solely as prompts for deliberative help seeking. When encountered during a goal-directed search process, messages highlighting harm or legal consequences may be especially effective at capturing attention and prompting disengagement from the immediate search trajectory. However, differences between active message variants were modest overall, and all active messages increased engagement relative to the neutral condition.  

The observed patterns in terms of message valence or framing also contrasts with expectations derived from behavior change and help-seeking research, which suggest that gain-framed messages emphasizing behavioral control, relief from distress, and accessible support may be more effective for promoting help resource engagement (e.g., \citealp{Gallagher2011, Hammer2024, Levenson2019}). However, differences between positively and negatively framed messages were modest, with overlapping confidence intervals, and should be interpreted with caution.

From a message framing perspective, the relative effectiveness of gain- and loss-framed appeals depends on how the advocated behavior is construed, particularly in terms of its perceived risk \citep{RothmanSalovey1997}. Although help seeking is often conceptualized as a prevention-oriented behavior, meta-analytic evidence suggests that framing effects are small and less consistent for higher-risk or detection-like actions \citep{Gallagher2011}. In the present context, engaging with support resources may be experienced as psychologically and socially risky, involving stigma, uncertainty, and potential negative consequences. Under such conditions, positive or gain framing of messages may not confer a clear advantage, and may even be marginally less effective than negative or loss-framed messages as appeared to be the case in our findings. 

More recent work further emphasizes that framing effects depend on the match between message, person, and context \citep{ROTHMAN202043}. It is therefore plausible that optimal framing varies across environments: individuals encountering warnings earlier in a trajectory of behavior (e.g., via clear web search) may be more responsive to gain-framed messages, whereas those engaging in more goal-directed or sustained search behavior (e.g., via dark web search engines) may respond more strongly to loss-framed messages. However, environment should not be treated as a direct proxy for trajectory. Although clear web search may often capture earlier-stage engagement, and dark web environments more sustained or intentional behavior, these associations are imperfect and likely to vary across users. For instance, younger users or those with greater technical familiarity may access dark web tools earlier in their trajectory. Survey findings suggest that individuals seeking CSAM on Ahmia are typically young, with 45\% in the 18-24 age range, and a substantial number also under 18 \citep{protectchildren2026csam}. Trajectory-based interpretation is necessarily tentative, but provides one potential account of the observed pattern.

Taken together, these findings suggest that message content matters, but that differences between specific framings are modest and context-dependent. All active messages increased engagement relative to a neutral baseline, supporting an approach in which platforms monitor and iteratively refine message content in response to user behavior and context, rather than relying on a single optimal content theme and framing. 

\subsection*{Future directions}

A limitation of a large-scale study of dark web search behavior such as this one is the inability to incorporate individual-level data, including demographic, psychological, attitudinal, or behavioral characteristics. This constrains our ability to examine how message effectiveness varies across individuals, despite theoretical and empirical indications that responses to warning messages are likely to depend on the interaction between message content, user characteristics, and behavioral context. Future work integrating experimental message variation with richer user-level or situational data may help to identify whether specific message types are differentially effective for particular subgroups or stages of offence trajectory. This is likely easiest to undertake on the clear web, where service providers often have additional information about the user. In addition, extending outcome measures beyond click-through rates to include indicators such as time spent on help resources or engagement with specific activities may provide further insight into whether different message types are associated with more sustained or meaningful help seeking.

Ahmia.fi occupies a particular position within the broader online ecosystem. It indexes content on the dark web---a key infrastructure for CSAM distribution \citep{Gannon2023}---and a substantial proportion of searches on the platform are flagged as potentially CSAM-seeking (\DescrTriggeringRateOverall{} during this study period). At the same time, it has longstanding measures in place to prevent indexing of harmful content and to promote access to help resources and prevention initiatives. As a result, the effectiveness of specific warning messages in this context may not generalize directly to other environments, including clear web search engines, social media platforms, messaging services, encrypted platforms, or other dark web spaces. Future work should therefore examine how message content performs across different technological and behavioral contexts.

The present study examined messages designed around distinct thematic categories. However, it remains unclear whether combining message elements (e.g., emphasizing harm alongside self-efficacy in seeking help) may yield additional gains. Future research could also explore complementary intervention levers, including visual design features \citep{prichard2024effect} and interactive approaches such as chatbots \citep{scanlan2026}. The increasing availability of scalable AI-based tools presents further opportunities to support both the delivery of help resources and the iterative refinement of messaging in response to user behavior at the platform or session level.

Finally, repeated exposure to warning messages introduces the potential for habituation, whereby users become less responsive over time \citep{amran2018habituation, kim2009habituation}. This study has clearly demonstrated that changing messages is achievable for platforms that are motivated to do so. The impact of changing messages, or the use of dynamic messaging, could help mitigate habituation and further increase help seeking \citep{watters2026}. Developing an evidence base that identifies not only which messages are effective, but also how their impact is sustained over repeated exposures, will be important for maintaining intervention effectiveness. Continued accumulation of evidence on how to close the gap between high-risk online activity and engagement with therapeutic resources is critical to inform innovation in this space.

\section*{Conclusion}

This study provides large-scale field evidence that the content of warning messages meaningfully shapes engagement with help-seeking resources among individuals actively searching for CSAM-related material. Across more than three million warning exposures in a live, high-anonymity environment, relatively small changes in message content produced consistent and measurable differences in behavior.

A key contribution of this work lies in its ecological validity. Rather than relying on hypothetical scenarios, self-report judgments, or simulated environments, the present study captures behavioral responses in situ, at the point of risk. The consistency of findings across randomized-day comparisons, campaign-wide analyses, and interrupted time series models strengthens confidence that observed effects reflect genuine responses to message content rather than artifacts of timing or external conditions.

The study also benefits from a strong pre-intervention comparison period. The 70-day pre-campaign phase involved the same detection system and an existing warning with a help-seeking pathway, allowing changes observed during the campaign to be interpreted as arising from message redesign rather than the introduction of warning infrastructure per se. This strengthens the inference that observed increases in engagement reflect differences in message content and presentation, rather than the mere presence of a warning.

The findings also have implications beyond the specific context of a dark web search engine. Although Ahmia.fi occupies a distinct position within the online ecosystem, the underlying mechanism—interrupting a goal-directed search and offering an immediate pathway to support—is shared across many platforms, including clear web search engines, social media, messaging platforms and other digital services. The present results therefore provide evidence that can inform intervention design more broadly, particularly in contexts where real-time behavioral disruption and redirection are feasible. For an exploration of whole system recommendations for governments, regulators, industry and civil society organizations in an accompanying report based on these findings, see \citealp{protectchildren2026preventive}. 

Importantly, the results do not support a simple “one best message” interpretation. While harm-focused messages showed the strongest effects in this setting, all active messages increased engagement relative to the neutral comparator used during the campaign, and differences between message types were modest. This reinforces the view that warning messages should be understood as part of an adaptive intervention system rather than a static solution. Their effectiveness is likely to depend on context, user characteristics, and stage of behavior, and may change over time with repeated exposure.

Taken together, these findings support a prevention-oriented approach in which warning messages function not only as deterrents, but as scalable gateways to support. In high-risk digital environments, where opportunities for intervention are brief, optimizing these moments may offer a practical means of increasing help seeking at scale. Continued progress will depend on iterative testing, cross-platform replication, and the integration of messaging within broader systems of support, with the aim of narrowing the gap between high-risk behavior and engagement with intervention.

%% file: sections/05_declarations.tex
\section*{Declarations}

\subsection*{Funding}
This report was produced as part of a project led by Protect Children and funded by the Home Office. Part of Caoilte Ó Ciardha's contribution to the study was funded by the Tech Coalition Safe Online Research Fund (Grant No. 23-EVAC-0015.2-University of Kent). Part of Juha Nurmi's contribution to the study was funded by the European Commission under the Horizon Europe funding programme, as part of the project SafeHorizon (Grant Agreement 101168562).